\documentclass[dvips]{article}

\newcommand{\degr}{\mbox{$^\circ$}}%

\usepackage{icrc2011}

\usepackage{url}

\title{Development of Non-sequential Ray-tracing Software for Cosmic-ray Telescopes}

\newcommand{\etal}{\MakeLowercase{\textit{et al. }}} 
\shorttitle{A.~Okumura \etal Development of Non-sequential Ray-tracing Software}

\authors{Akira Okumura$^{1,2}$, Masaaki Hayashida$^{2, 3}$, Hideaki Katagiri$^{4}$, Takayuki Saito$^{5}$, Vladimir Vassiliev$^{6}$}
\afiliations{
$^1$Institute of Space and Astronautical Science, JAXA, Sagamihara, Kanagawa 252-5210, Japan\\
$^2$Kavli Institute for Particle Astrophysics and Cosmology, SLAC National Accelerator Laboratory, Menlo Park, CA 94025, USA
\\
$^3$Department of Astronomy, Graduate School of Science, Kyoto University, Sakyo-ku, Kyoto 606-8502, Japan\\
$^4$College of Science, Ibaraki University, Mito 310-8512, Japan
\\
$^5$Max-Planck-Institut f\"{u}r Physik, F\"{o}hringer Ring 6, 80809 M\"{u}nchen, Germany\\
$^6$Department of Physics and Astronomy, University of California Los Angeles, CA 90095-1547, USA}
\email{oxon@astro.isas.jaxa.jp (Akira Okumura)}

\abstract{We have developed non-sequential ray-tracing software which is aimed to be widely used, along with air-shower simulations, in the design of optical systems for cosmic-ray experiments. The code is based on the ROOT geometry library to provide a non-sequential photon tracking system, which is valuable when simulating refraction and multiple reflections. In addition to the basic ROOT classes, we have implemented new geometry ROOT classes so that users can flexibly define various geometries such as aspherical or Winston-cone type surfaces. We demonstrate the capabilities and performance of the software with examples of optical systems used in current and future experiments.}

\begin{document}
\maketitle

\section{Introduction}
\quad Optical systems for modern cosmic-ray (CR) experiments play an important role to improve their detection sensitivities of very-high-energy (VHE) gamma rays or ultra-high-energy (UHE) CRs. The basic strategy in the design of an optical system is to simultaneously achieve a large effective area of mirrors ($\sim5\ \mathrm{m}^2$ to $\sim500\ \mathrm{m}^2$), a wide field of view ($\sim5\degr$ to $\sim50\degr$ in diameter), a high angular resolution ($\sim0.1\degr$ to $\sim1\degr$), and low cost per telescope. In order to satisfy these requirements that are not demanded for optical-to-infrared telescopes, some techniques such as segmented mirrors and light guides are widely used for CR telescopes. Consequently optical simulations of CR telescopes become more puzzling, as their optical systems get more complicated.

\quad Ray-tracing software is often developed independently in each CR experiment or institute. This is because existing programs for optical telescopes are not adequate for CR telescopes (see details in Subsection~\ref{subsec_requirements}). However, many requirements in optical simulations are common in all CR experiments. This means that development of a convenient and accurate ray-tracing program shared by the CR community will enables us to save the time for software development and debugging. Our objective in the present paper is to offer a standard ray-tracing program to be used quite easily for many optical studies in CR experiments, and to reduce dispensable effort in parallel software development.

\section{ROBAST}
\label{sec_ROBAST}
\quad We have developed a new ray-tracing program, ROBAST\footnote{\url{http://sourceforge.net/projects/robast/}} (ROot-BAsed Simulator for ray Tracing), which is based on the ROOT framework \cite{Brun:1997:ROOT----An-object-oriented-data-analysis}. ROBAST fully utilizes its geometry library \cite{Brun:2003:The-ROOT-geometry-package} to offer several functionalities required for ray-tracing simulations of CR telescopes.

\subsection{Requirements and Functionalities}
\label{subsec_requirements}

\subsubsection{Non-sequential Ray Tracing}
\quad The normal ray-tracing method adopted in most programs is so-called ``sequential ray tracing'' in which the order of all surfaces in an optical system is defined by the user, and photons reach the surfaces sequentially. It is often used for simulations of simple optical systems, such as a single-dish parabolic mirror. On the contrary, we need ``non-sequential ray-tracing'' mode to simulate scattered photons or multiple segmented mirrors. This is because the order of multiple reflections and refraction cannot be specified uniquely by the user.

\begin{figure}[tb]
  \vspace{5mm}
  \centering
  \includegraphics[width=3.2in]{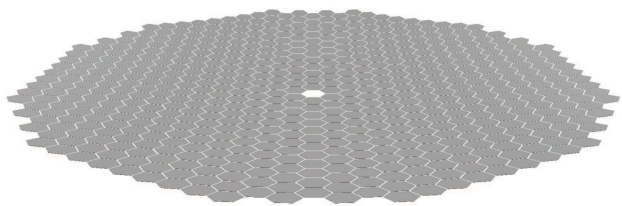}
  \caption{An example of hexagonal segmented mirrors aligned on a parabolic surface. The same optical system will be used for the large size telescopes in the Cherenkov Telescope Array project \cite{The-CTA-Consortium:2010:Design-Concepts-for-the-Cherenkov-Telesc}.}
  \label{fig_LST}
\end{figure}

\quad Segmented mirrors are almost always used in CR experiments to enlarge the effective mirror area of each telescope, and to reduce the total cost per area. Each mirror segment is usually placed on a parabolic or spherical surface, or aligned on the Davies-Cotton configuration as shown in Figure~\ref{fig_LST} \cite{Davies:1957:Design-of-the-Quartermaster-So,Weeks:1989:Observation-of-TeV-Gamma-Rays-,Kawachi:2001:The-optical-reflector-system-f,Bernlohr:2003:The-optical-system-of-the-H.E.S.S.-imagi,Holder:2006:The-first-VERITAS-telescope,Oliveira:2004:Manufacturing-the-Schmidt-corr}. These mirrors cannot be simulated by sequential ray-tracing software, because the intersection point of an input photon with each mirror segment must be calculated. The non-sequential mode is also required when we simulate light guides such as Winston cones \cite{Winston:1970:Light-Collection-within-the-Fr} or similar surfaces, because photons can be reflected by light guides multiple times \cite{Kabuki:2003:Development-of-an-atmospheric-Cherenkov-,Bernlohr:2003:The-optical-system-of-the-H.E.S.S.-imagi}.

\quad Coding a non-sequential ray-tracing algorithm from scratch is not simple, but it has been already established as particle tracking systems in high-energy physics. Since powerful particle tracking systems are already available in ROOT \cite{Brun:2003:The-ROOT-geometry-package} and Geant4 \cite{Agostinelli:2003:Geant4--a-simulation-toolkit}, we have decided to use the ROOT geometry library as the ray tracer in ROBAST. The library also enables us to satisfy other requirements.

\subsubsection{Flexibility of Geometry Construction}

\begin{figure}[tb]
  \vspace{5mm}
  \centering
  \includegraphics[width=3.2in]{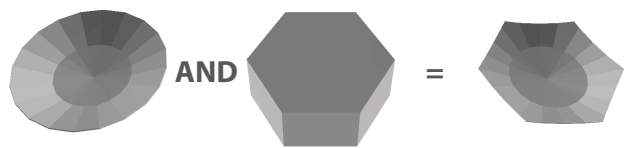}
  \caption{An example of a mirror segment built with a ``boolean'' operation of a partial sphere and a hexagonal cylinder. The region spatially shared with two objects  is used as a new object.}
  \label{fig_AND}
\end{figure}

\quad As shown in Figure~\ref{fig_LST}, hexagonal or rectangle segmented mirrors are often used \cite{Holder:2006:The-first-VERITAS-telescope,Oliveira:2004:Manufacturing-the-Schmidt-corr} as well as circular mirrors \cite{Kabuki:2003:Development-of-an-atmospheric-Cherenkov-,Bernlohr:2003:The-optical-system-of-the-H.E.S.S.-imagi}. Each segment shape can be easily built using basic geometry shapes\footnote{See e.g. \url{http://root.cern.ch/root/htmldoc/TGeoBBox.html} and derived classes.} and ``boolean'' operations offered by ROOT. For example, a hexagonal mirror can be defined by an ``AND'' operation of a sphere and a hexagonal cylinder as illustrated in Figure~\ref{fig_AND}. In addition to mirror shapes, other optical components and telescope structures can be built. This functionality is important for accurate calculations of effective areas and shadowing effects of telescopes.

\begin{figure}[tb]
  \vspace{5mm}
  \centering
  \includegraphics[width=2.2in]{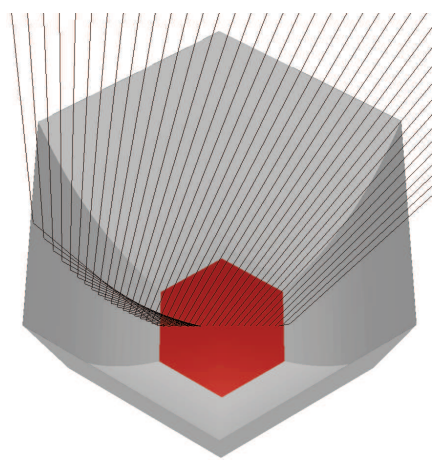}
  \caption{An example of a hexagonal Winston cone built with \texttt{AGeoWinstonConePoly} class consisting of six inclined parabolas. The input and output radii are 20~mm and 10~mm, respectively. Photon tracks simulated by \texttt{ARay} class instances are shown with the black lines. The red hexagon imitates the photocathode of a photomultiplier tube (PMT).}
  \label{fig_WinstonCone}
\end{figure}

\begin{figure}[tb]
  \vspace{5mm}
  \centering
  \includegraphics[width=3.in]{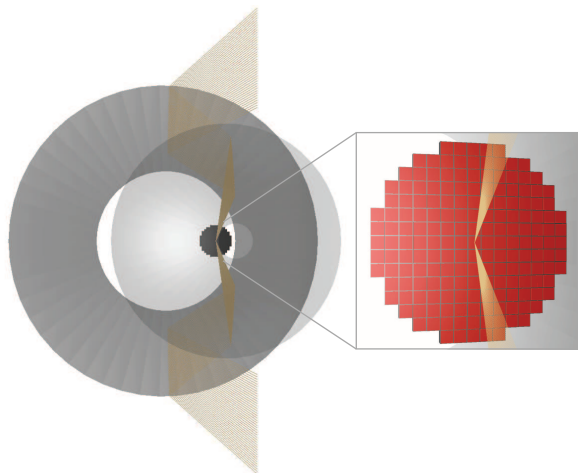}
  \caption{The Schwarzschild-Couder optical system \cite{Vassiliev:2007:Wide-field-aplanatic-two-mirro} built with \texttt{AGeoAsphericDisk} class. The primary (larger doughnut) and the secondary (semitransparent smaller doughnut) mirrors are approximated by 14th-order polynomials. 177 multi-anode PMTs (red squares) are spherically aligned on the focal surface. The diameters of the primary mirror, the secondary mirror, and the focal plane are, 9.7~m, 5.4~m, and 0.78~m, respectively.}
  \label{fig_SC}
\end{figure}

\quad The shape of a hexagonal Winston cone or a similar light guide is mathematically expressed by a combination of simple geometries, such as paraboloids and cones. However, due to the limited numbers of ROOT geometry shapes, even a basic hexagonal Winston cone cannot be built with the default ROOT classes. In addition, some optical systems recently designed use more complicated shapes that are not available in ROOT \cite{Sasaki:2002:Design-of-UHECR-telescope-with,Vassiliev:2007:Wide-field-aplanatic-two-mirro}. We have added two more geometry classes, \texttt{AGeoWinstonConePoly} and \texttt{AGeoAsphericDisk}, in ROBAST so that users can build complex geometries as demonstrated in Figures~\ref{fig_WinstonCone} and \ref{fig_SC}. Of course new geometry classes can be easily implemented by the user to build more complicated shapes.

\quad It is also easy for a user to place an optical component at any coordinates with any angles using translation and rotation matrices. This feature is important when the user evaluates the degradation of optical performance caused by misalignment of optical components.

\subsubsection{Visualization}
\quad As already shown in Figures~\ref{fig_LST} to \ref{fig_SC}, ROBAST is able to visualize geometries thanks to the ROOT OpenGL viewer. Visualization plays a major role when user-defined geometries are inspected by eye. However, visualization functionality has not been well integrated for self-made programs because it is difficult to develop a new visualization system from scratch. The built-in visualization engine will also be helpful for presentations.

\subsubsection{Connectivity with Other Software}
\quad ROBAST has been developed especially for CR telescope simulations. It depends on other external software. For example, ray-tracing software is frequently used with air-shower Monte Carlo simulations or electronics simulations in numerous batch jobs. In addition, ROBAST results are expected to be analyzed seamlessly by standard tools such as ROOT (CINT or C++) or Python libraries. Since ROBAST runs in either compiled C++ code, ROOT CINT scripts, or Python scripts, most CR researchers can immediately use it with other external software. On the other hand, existing ray-tracing programs will not generally work this way.

\subsubsection{Multi Platform Support}
\quad ROBAST runs on all of the three major operation systems (OSs) used by high-energy astrophysicists: Mac OS X, Linux, and Windows.

\subsection{Comparison with Other Programs}



\quad In the previous subsection, we listed requirements for ray-tracing software and described the advantages that come with ROBAST. In this subsection, we compare ROBAST with other major programs, and describe some strengths and weaknesses.

\subsubsection{Geant4}
\quad Geant4 is a simulation toolkit for particle tracking and interaction \cite{Agostinelli:2003:Geant4--a-simulation-toolkit}. It is able to track optical photons as well as X-rays and gamma rays so that a user can simulate scintillation detectors. The optical photon tracking in Geant4 is similar to that of ROBAST, while it has two main advantages which are not currently available in ROBAST.

\quad The first advantage is that Geant4 is able to simulate the roughness of optical boundaries, and to choose the reflection angle distribution from a few options. Using these capabilities, a user can build more realistic optical components, e.g. a crystal scintillator wrapped in a diffusive reflector. Since this functionality is critical to simulate the mirror roughness, it will be implemented in the next release of ROBAST.

\quad The other is that Geant4 can import very complicated volumes defined in a GDML \cite{Chytracek:2006:Geometry-Description-Markup-Language-for} file along with \texttt{G4TessellatedSolid} class which consists of multiple polygon facets. ROOT and ROBAST are also able to import GDML files, but they do not have an equivalent of \texttt{G4TessellatedSolid}. Therefore, Geant4 is preferable if a user needs to convert his/her complicated CAD file into GDML.

\quad If a user is more familiar with Geant4 than ROOT, the user can select it for ray-tracing simulation. But it does not have similar classes as \texttt{AGeoAsphericDisk} and \texttt{AGeoWinstonConePoly} which are quite important for CR telescopes. In addition, ROOT is more widely used than Geant4.

\subsubsection{ZEMAX}
\quad There are many commercial or open source ray-tracing programs which are not being developed especially for CR telescopes. A commercial software, ZEMAX\footnote{\url{http://www.zemax.com/}} which runs on Windows only, is one of the most famous programs used in many aspects of optical simulations. The advantages of the sophisticated commercial software are as follows.

\quad First, it has a built-in optimization engine. The surface of optical components can be very quickly optimized from its graphical interface, especially in the non-sequential mode. But ROBAST does not have any optimization engine specifically designed for optical systems.

\quad Second, ZEMAX has many built-in analysis tools. However, optical systems used in CR experiments are less precise than other systems. For example, our optical systems do not reach the diffraction limit, and thus, we do not need high performance tools in most situations.

\quad ZEMAX or similar software would be better than ROABST at the first stage of designing optical systems. However, ROBAST will suit all our requirements after optimizations. An advanced ROBAST user can build own optimizer using MINUIT distributed with ROOT.

\section{Simulation Examples}
\begin{figure}[tb]
  \vspace{5mm}
  \centering
  \includegraphics[width=3.2in]{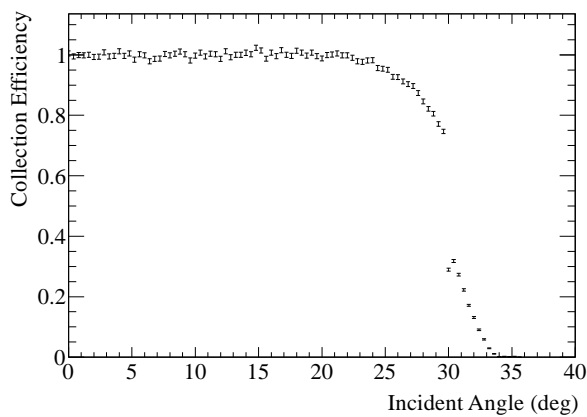}
  \caption{The collection efficiency of the hexagonal Winston cone shown in Figure~\ref{fig_WinstonCone}. $\sim14,000$ random photons were simulated for each incident angle. The mirror reflectivity was assumed to be 100\%. The discontinuity at $\sim30^\circ$ is due to the two parabolic surfaces vertical to the incident photons in the figure.}
  \label{fig_WinstonCone2}
\end{figure}

\quad Figure~\ref{fig_WinstonCone2} is a simulation example demonstrating the collection efficiency of the Winston cone shown in Figure~\ref{fig_WinstonCone}. In producing this simulation, the photon tracking is transparent to the user who only has to build the cone's shape in a ROOT script.

\begin{figure}[tb]
  \vspace{5mm}
  \centering
  \includegraphics[width=3.2in]{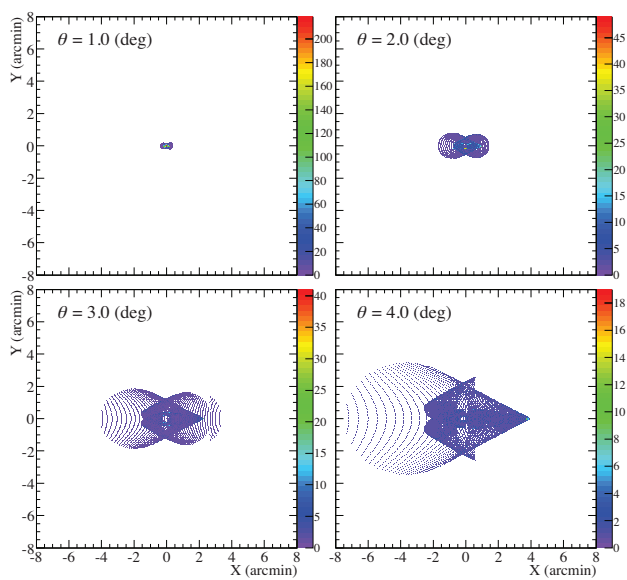}
  \caption{Spot diagrams of the Schwarzschild-Couder optical system shown in Figure~\ref{fig_SC}.}
  \label{fig_spot}
\end{figure}

\quad Figure~\ref{fig_spot} shows the spot diagrams of the Schwarzschild-Couder optical system shown in Figure~\ref{fig_SC}. The photon coordinates and directions on the focal plane are available within the same ROOT script that produced this simulation example. It is easy to produce these spot diagrams and the incident angle distribution on the focal plane on the fly with the ROOT analysis classes.

\section{Conclusion}
\quad We have developed a non-sequential ray-tracing program which can be used in many aspects of CR experiments. ROBAST satisfies many requirements for studies of CR telescopes. It is already being used for some optical studies for the CTA and the MAGIC experiments. We expect ROBAST to be also used in other experiments in near future.

\section{Acknowledgements}
\quad We are grateful to Dr. Andrei Gheata who is one of the maintainers of the ROOT geometry library. He kindly answered many questions at the first stage of the ROBAST development. We could not start this project without the ROOT geometry library and his great support. We would also like to thank Dr. Akito Kusaka who gave A.~O. an initial idea of ray-tracing code. A.~O. and M.~H. are supported by Grant-in-Aid for JSPS fellows.

\end{document}